\documentclass[twocolumn, linenumber]{aastex62}

\usepackage{amsmath}
\usepackage{array}
\usepackage{natbib}
\usepackage{color}
\turnoffedit
\newcommand{\mytilde}{\raise.19ex\hbox{$\scriptstyle\sim$}}

\shorttitle{Weak-lensing Study of SpARCS1049}
\shortauthors{Finner et al. }

\begin{document}

\title{Constraining the Mass of the Emerging Galaxy Cluster S\lowercase{p}ARCS1049+56 at $\lowercase{z}= 1.71$ with Infrared Weak Lensing}

\correspondingauthor{M. James Jee}
\email{mkjee@yonsei.ac.kr}
\email{kylefinner@gmail.com}
\author{Kyle Finner}
\affil{Department of Astronomy, Yonsei University, 50 Yonsei-ro, Seoul 03722, Korea}

\author{M. James Jee}
\affil{Department of Astronomy, Yonsei University, 50 Yonsei-ro, Seoul 03722, Korea}
\affil{Department of Physics, University of California, Davis, One Shields Avenue, Davis, CA 95616, USA}

\author{Tracy Webb}
\affil{Department of Physics, McGill University, 3600 rue University, Montreal, Quebec, H3P 1T3, Canada}
\author{Gillian Wilson}
\affil{Department of Physics and Astronomy, University of California Riverside, 900 University Avenue, Riverside, CA 92521, USA}
\author{Saul Perlmutter}
\affil{Physics Division, Lawrence Berkeley National Laboratory, 1 Cyclotron Road, Berkeley, CA, 94720, USA}
\author{Adam Muzzin}
\affil{Department of Physics and Astronomy, York University, 4700 Keele St., Toronto, Ontario,MJ3 1P3, Canada}
\author{Julie Hlavacek-Larrondo}
\affil{Department of Physics, Universite de Montreal, Montrea, Quebec H3T 1J4, Canada}
\begin{abstract}

In the hierarchical structure formation model of the universe, galaxy clusters are assembled through a series of mergers. Accordingly, it is expected that galaxy clusters in the early universe are actively forming and dynamically young. Located at a high redshift of $z=1.71$, SpARCS1049+56 offers a unique look into the galaxy cluster formation process. This cluster has been shown to be rich in cluster galaxies and to have intense star formation\deleted{coincident with the brightest cluster galaxy}. Its high redshift pushes a weak-lensing analysis beyond the regime of the optical spectrum into that of the infrared. Equipped with deep \textit{Hubble Space Telescope} Wide Field Camera 3 UVIS and IR observations, we present a weak-lensing characterization of SpARCS1049+56. As few IR weak-lensing studies have been performed, we discuss the details of PSF modeling and galaxy shape measurement for an IR weak-lensing procedure and the systematics that come with the territory. It will be critical to understand these systematics in future weak-lensing studies in the IR with the next generation space telescopes such as JWST, Euclid, and WFIRST. Through a careful analysis, the mass distribution of this young galaxy cluster is mapped and \replaced{shown to have a relaxed morphology}{the convergence peak is detected at a 3.3$\sigma$ level}. The weak-lensing mass of the cluster is estimated to be $3.5\pm1.2\times10^{14}\ \text{M}_\odot$ and is consistent with the mass derived from a mass-richness scaling relation. This mass is extreme for a cluster at such a high redshift and suggests that SpARCS1049+56 is rare \replaced{for our}{in the standard} $\Lambda$CDM universe.

\end{abstract}

\keywords{
gravitational lensing ---
dark matter ---
cosmology: observations ---
X-rays: galaxies: clusters ---
galaxies: clusters: individual (SpARCS1049+56) ---
galaxies: high-redshift}

\section{Introduction}
Galaxy clusters emerged from the largest overdensities in the primordial universe. Their evolution is sensitive to both the growth rate of structure and the expansion history of the universe. For this reason, they are a useful probe to test cosmological theories. The observed size, mass, and abundance of galaxy clusters are a valuable tool to constrain the parameters that formulate our cosmological models. In particular, the abundance of galaxy clusters is a sensitive probe of the matter density $\Omega_\text{m}$ and the normalization of the matter power spectrum $\sigma_8$ \citep[e.g.][]{2007gladders}. However, the strong degeneracy between $\sigma_8$ and $\Omega_\text{m}$ prevents the constraint of each parameter independently with the cluster mass function alone. This degeneracy can be alleviated by combining cluster mass functions over a wide range of redshift \citep[e.g.][]{2006albrecht}. 

A dominant systematic uncertainty in using galaxy clusters as cosmological probes is their mass calibration. Many of the large studies of galaxy clusters estimate mass through scaling relations, such as velocity dispersion or X-ray temperature, that rely on equilibrium or quasi-equilibrium state assumptions. The systematic errors innate to the mass estimate are then inherited by the cosmological constraint. Weak lensing (WL hereafter) provides a mass estimation free of an assumption of the dynamical state of the cluster and has the ability to provide more robust mass estimates. This merit is particularly important for galaxy clusters at high redshift where they tend to be in an early stage of formation and thus subject to a large departure from dynamical equilibrium.

To date, very few high-redshift galaxy clusters have been measured with WL. The vast majority of WL surveys have been focused on redshift less than unity. In fact, of the large WL surveys, the \textit{Hubble Space Telescope} (\textit{HST} hereafter) studies of \cite{2011jee} and \cite{2018schrabback} are the only to include clusters at $z>1$. Beyond a redshift of 1.5, only a single galaxy cluster has been studied with WL, IDCS J1426+3508 \citep{2016mo, 2017jee} at redshift 1.75. The lack of studies at high redshift can primarily be attributed to the difficulty of detecting the lensing signal. The lensing distortions are caused by a massive intervening object between source galaxies and the observer. When a high-z cluster is the lens, more distant galaxies need to be probed to detect the lensing signal. This requires very deep imaging \added{at infrared wavelengths} to robustly detect galaxies in the 25-28th magnitude range. Fortunately, \added{some imaging programs with} the \textit{HST} \replaced{is}{are} probing these depths of the universe. 

One of the goals of the See Change program (PI: Perlmutter) is to probe the WL mass function of galaxy clusters at redshift greater than one. The See Change sample includes 11 galaxy clusters in the redshift range $1.10$ to $1.75$, with IDCS J1426+3508 the highest. The second highest redshift cluster in the sample is SpARCS1049+56 (hereafter SpARCS1049 for brevity) and it is the focus of this study. 

SpARCS1049 was \deleted{first}discovered in the \textit{Spitzer} Adaptation of the Red-sequence Cluster Survey (SpARCS) \citep{2009muzzin, 2009wilson}. This survey utilized a two IR filter system to detect galaxy overdensities by the 4000\AA\ break \citep{2006wilson}. The survey footprint included 11 square degrees of the Lockman Hole, a 59 square degree region that is relatively clear of galactic HI emission and within this region lies SpARCS1049. % at 10h49m22.6s +56d40m32.5s 

The first detailed study of SpARCS1049 was achieved by \cite{2015webb}. They used the archival \textit{Spitzer} observations and supplemented them with their own observations of the cluster from the James Clark Maxwell Telescope, \textit{HST}, and Keck. Their Keck-MOSFIRE spectroscopy determined the galaxy overdensity redshift to be centered at $z=1.709$. Based on this redshift, they classified \added{27} cluster member galaxies \replaced{to be}{as} those within $1500$ km s$^{-1}$ and 1.8 Mpc projected distance of the brightest cluster galaxy (BCG). The velocity dispersion ($\sigma=430^{+80}_{-100}$ km s$^{-1}$) of these galaxies provides a mass M$_{vir}$ of $8\pm3\times10^{13}\ \text{M}_\odot$. The authors go into detail about the shortcomings of the velocity dispersion from this sample. Their classification of cluster member galaxies goes well beyond the expected virial radius of the cluster (\mytilde1 Mpc). Futhermore, the redshifts were detected by the H$\alpha$ emission line, which only selects active galaxies. In addition to this mass estimate, they found the richness of the cluster to be $N_{\text{gal}}=30\pm8$ and used the mass-richness scaling relation from \cite{2014andreon} to infer a mass M$_{500\text{kpc}}$ of $3.8\pm1.2\times10^{14}\ \text{M}_\odot$.   

We present a WL characterization of SpARCS1049 through the \textit{HST} Wide Field Camera 3 (WFC3) IR filters. The mass estimate from WL is an independent test of the previous two masses because it does not rely on the dynamical state of the galaxy cluster. 
WL using the \textit{HST} IR filters has been achieved once before in \cite{2017jee}. Their WL analysis of SPT-CL J2040-4451 (z=1.48) and IDCS J1426+3508 (z=1.75) clearly detected the WL signals and quantified the masses of the two young, massive clusters.

In \textsection\ref{section:data_reduction} we describe the \textit{HST}-IR observations, data reduction, and PSF modeling. The details of WL and our shape measurement pipeline are outlined in \textsection\ref{sec:wl_method}. We present our mass map and mass estimation in \textsection\ref{sec:results}. The mass of the cluster and its rarity \replaced{is}{are} discussed in \textsection\ref{sec:discussion} before we conclude in \textsection\ref{sec:conclusion}.

In this paper, we use the cosmological parameters from \cite{2016planck}. The notation M$_{200}$ represents a spherical mass within the radius $r_{200}$, inside which the mean density is equal to 200 times the critical density of the universe at the cluster redshift. At $z= 1.71$, the plate scale is \mytilde8.70 kpc$/\arcsec$.    

\section{Observations} \label{section:data_reduction}
Observations of SpARCS1049 were obtained with the \textit{HST} in programs 13677 (PI: S.~Perlmutter) and 13747 (PI: T.~Webb) \added{\edit1{from 2014 February to 2015 May}}. In both programs the cluster was imaged with WFC3 using the UVIS F814W and the IR F105W/F160W filters. Combining the two programs, the total exposure times are 2846s, 8543s, and 9237s for F814W, F105W, and F160W, respectively. The joining of these two programs provides very deep imaging data, which is critical for resolving faint source galaxies in high-$z$ cluster WL. 
Both observing runs were centered on the BCG location with camera rotations and small dithers between pointings. This technique is ideal for WL analyses because it minimizes the effect of diffraction spikes in stacked images and improves sampling of the point spread function (PSF). For our WL analysis, we use the F160W coadd to measure shapes because it is the deepest among the three filters and also the emission in this bandpass represents the rest-frame optical emission of source galaxies at $z\sim2$, which has a smoother light profile than bluer light that traces clumpy star formation regions of high redshift galaxies.

The calibrated individual exposures (FLT/FLC images) were retrieved from the Mikulski Archive for Space Telescopes (MAST)\footnote{https://archive.stsci.edu/}. Prior to retrieval, these exposures were processed by the STSci OPUS pipeline using the \textit{calwf3} software task. The \textit{calwf3} task performs the standard calibration steps of dark subtraction, flat fielding, etc. Note that the calibration methods for the WFC3-UVIS and WFC3-IR detectors differ in some aspects. The WFC3-UVIS channel is a CCD detector and has a degraded ability to transfer charges during readout. Recent versions of the \textit{calwf3} task correct for charge transfer efficiency (CTE) \added{degradation} \citep{2016bajaj}. On the other hand, the WFC3-IR detector does not perform readout through charge transfer as CCDs do and thus does not suffer from CTE \added{degradation}. However, the detector possesses other systematic effects, which we discuss in \textsection\ref{sec:systematics}. 

\texttt{Multidrizzle} \citep{2003koekemoer} was used on the calibrated exposures to perform cosmic ray rejection, sky subtraction, and geometric distortion correction. Individual exposures were ``single-drizzled" to a north-up orientation with the common World Coordinate System (WCS) to prepare them to be stacked into a mosaic image. 
We then performed alignment of the individual exposures by iterative minimization of the offset of astronomical sources that are common within overlapping regions. This method of alignment was shown to be sufficient for cluster WL applications in \cite{2014jee}. With the astrometric solution obtained, a second \texttt{Multidrizzle} was performed to combine the images into a well-aligned, stacked mosaic. 

\begin{figure*}[!ht]
    \centering
    \includegraphics[width=\textwidth]{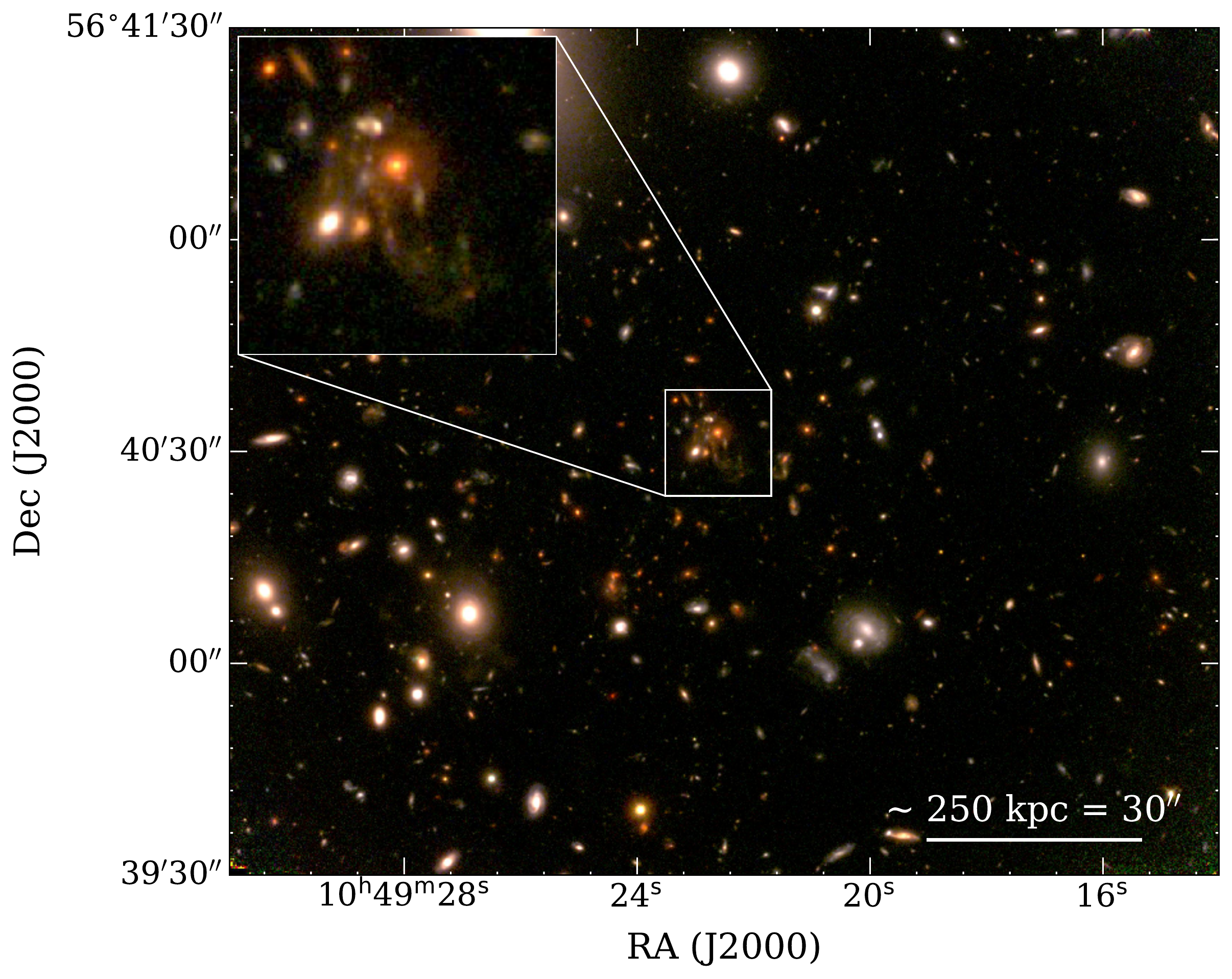}
    \caption{\textit{HST} color-composite image of SpARCS1049 from stacking the F160W, F105W, and F814W filter images as RGB, respectively. The deep orange galaxy at the center of the image is the BCG (10$^\text{h}$49$^\text{m}$22$^\text{s}$.6, 56$^\circ$40\arcmin33\arcsec) and is shown in the inset image. The magnificent tidal feature discussed in \cite{2015webb} is seen in the inset image stretching from the center to the southwest.}
    \label{fig:color_image}
\end{figure*}

We chose to tune the input parameters of \texttt{Multidrizzle} to optimize the F160W image quality as it is used for our lensing analysis. The full width at half maximum (FWHM) of the PSF in the IR detector is fractionally larger (FWHM\mytilde$0\farcs16$) than the \added{native} pixel scale 0\farcs13. This causes \deleted{an}undersampling of the PSF. The DrizzlePac handbook \citep{2012drizzle} suggests that upsampling to a final pixel scale that samples the PSF by about 2.0 to 2.5 pixels is ideal. Following this advice, we chose a final pixel scale of 0\farcs05 pix$^{-1}$ to \replaced{minimize}{mitigate} the effect of undersampling the PSF. Although this pixel scale is larger than the UVIS \added{native} pixel scale of 0\farcs04, the downsampled F814W images are strictly used for color image generation and not in the scientific analysis. We set \texttt{final\_pixfrac} to 0.7 and used a Gaussian kernel to drizzle the images. The color-composite image in Figure \ref{fig:color_image} was created by combining the F160W, F105W, and F814W filter images. The BCG is the deep orange galaxy located in the center of the image with the ``beads-on-a-string'' interacting galaxy stretching from east of the BCG to \mytilde 50 kpc southwest. These features are more obvious in the zoomed inset. For more on the galaxies of SpARCS1049 see \cite{2015webb}, \cite{2017webb}, and \cite{2019trudeau}.     

We created a detection image by weight-averaging the F105W and F160W images with weights from \texttt{Multidrizzle}. Objects were detected with \texttt{Source Extractor}\footnote{https://www.astromatic.net/software/sextractor} \added{\edit1{\citep{1996bertin}}} in dual-image mode by selecting sources in the detection image and measuring them in each filter-specific image (F105W or F106W). Objects that subtend at least 5 pixels having signal at least 1.5-$\sigma$ above the background \replaced{level}{rms} were measured. WL studies using the \textit{HST} have shown that the background galaxy density is high (\mytilde100 galaxies arcmin$^{-2}$). An issue that arises with high source galaxy density is blending (overlapping) of galaxy images. \cite{2018mandelbaum} discusses the bias arising from deblending galaxies in detail. The primary concern for deblending in this study occurs when two images overlap from galaxies at large separation in redshift. To mitigate the effect, we deblended objects using \texttt{Source Extractor} with \texttt{DEBLEND\_NTHRESH} = 8 and \texttt{DEBLEND\_MINCONT} = 0.005. However, such rigorous deblending can cause a foreground galaxy to be deblended into multiple objects. These spurious detections, that contain no lensing signal, were removed after source selection (Section \ref{section:source_selection}) through visual inspection. In total, \mytilde6,900 objects were detected in the $\mytilde 3\arcmin \times 3\arcmin$ WFC3-IR mosaic image and compiled into an object catalog.

\subsection{PSF Model}
Ground-based WL analyses rely heavily on the correction of the PSF as it causes a significant dilution of the observed lensing shear. In addition, the PSF tends to have a characteristic direction that mimics shearing. These two PSF effects are also present in the space-based \textit{HST} imaging but to a lesser extent because of the lack of atmosphere.

This is the first study to use the WFC3-IR/F160W channel for a WL analysis. We modeled the PSF using a version of our PSF modeling pipeline based on principal component analysis (PCA) and updated for the F160W channel. This pipeline has been described in detail in our previous papers \citep[e.g.][]{2007jee, 2017finner}. Here we will briefly explain the PSF pipeline for the WFC3-IR/F160W channel and refer the reader to our previous work for an in-depth discussion. 

A major hurdle for modeling the PSF of \textit{HST}, which depends on time and position on the focal plane, is the lack of stars available in a single science frame. Fortunately, the \textit{HST} PSF variation possesses a repeatable pattern \citep{2007jee} \added{\edit1{that is dependent on the focus (breathing) of the telescope following the 1.5 hour orbit of the telescope around Earth}}. This allows a utilization of dense archival stellar images to model the PSF, which can then be applied to the science frames that are taken at a different epoch.
Table \ref{table:psf_frames} contains a list of the dense stellar fields that we tested for our PSF modeling pipeline. In the majority of these fields, the frames are overcrowded with stars and overlapping diffraction spikes significantly hamper our ability to model the PSF. However, the exposures of NGC104 (also known as 47 Tuc) and NGC2808 in programs 11453, 11664, and 11665 contain the best spatial star sampling to characterize the PSF and we relied on these frames for our PSF pipeline. These images were drizzled with the same settings as the single-drizzled science images (Section \ref{section:data_reduction}) and will be referred to as the stellar frames from here on. We ran our PSF modeling pipeline on the stellar frames and designed a position-dependent PSF for each frame. 

Switching to the science frames, we selected several stars ($5\sim10$) from each single-drizzled science image and recorded their pixel coordinates and ellipticity. At the coordinates of these stars, we retrieved the modeled PSF for each stellar frame. This resulted in a catalog of PSFs for each stellar frame at the defined science frame's star locations. To find the best-fit model stellar frame, we minimized the difference between the ellipticity of the modeled PSFs and science stars. \added{\edit1{The median reduced $\chi^2$ value is 1.8 for the best-fit models. Furthermore the residual ellipticities when comparing our model to the measured stellar ellipticities are $de\mytilde0.008$, which is sufficient for cluster lensing.}} Finally, PSFs for all objects in the F160W mosaic image were built by retrieving the best-fit PSF model at each object location for each science frame and stacking them into a final PSF.

\begin{table}[]
\caption{Archived F160W \textit{HST}/WFC3-IR images tested for our PSF modeling pipeline. Bold-font programs were selected for PSF modeling.}
\label{table:psf_frames}
\centering
\def\arraystretch{1.5}
\begin{tabular}{l c c c}
\hline \hline
Object & Program ID & Exposures & \added{\edit1{Obs. year}} \\ \hline
\textbf{47 Tuc}      & \textbf{11453}    & \textbf{18} & \textbf{2009}             \\ 
\textbf{NGC104}      & \textbf{11664}       & \textbf{6} & \textbf{2010}            \\
\textbf{NGC2808}      & \textbf{11665}       & \textbf{6} & \textbf{2011}            \\ 
NGC6388      & 11739       & 10 & 2010              \\ 
NGC6441      & 11739       & 20 & 2010              \\ 
OmegaCen      & 11928       & 27 & 2009             \\ 
OmegaCen      & 12353       & 15 & 2011             \\ 
OmegaCen      & 13691       & 6 & 2015              \\ 
\hline \hline
\end{tabular}
\end{table}

\subsection{WFC3-IR Detector WL Systematics} \label{sec:systematics}
Systematic effects inherent to the IR detector are a cause for concern for WL studies because they may falsely contribute to the WL signal. In the first WL analysis to use the WFC3-IR detector, \cite{2017jee} reported four systematic effects that need to be considered: interpixel capacitance (IPC), persistence, detector non-linearity, and undersampling. Readers are referred to \cite{2017jee} for detailed discussions on these four topics from a WL perspective. Here, after briefly describing these aforementioned effects, we will provide a detailed discussion on the brighter-fatter effect.

\textbf{IPC:} The WFC3-IR detector is a 1024x1024 HgCdTe array with a plate scale of $0\farcs13$ per pixel. \cite{2006brown} investigated the correlated noise in HgCdTe detectors and found charge sharing between neighboring pixels from capacitive coupling. This IPC is also present in the WFC3-IR detector \citep{2011hilbert}. We follow the same method as \cite{2017jee} and let our PSF model correct for IPC.  

\textbf{Persistence:} IR detectors are also susceptible to a persistence of signal after a reset. The effect is described in detail in \cite{2008smith}. Their investigation showed that the persistence of charge is greater for pixels that have been exposed near saturation in previous imaging. The STScI provides a tool\footnote{https://archive.stsci.edu/prepds/persist/search.php} to search for persistence in archived observations. Our search shows that persistence levels are low in observations of SpARCS1049, with a persistence of $\gtrsim0.01$ e$^-$ s$^{-1}$ in at most 0.1\% of the pixels and $\gtrsim0.1$ e$^-$ s$^{-1}$ for 0.03\% of the pixels. 

\textbf{Undersampling:} The FWHM of the  WFC3-IR F160W PSF is approximately the same size as the native plate scale (0\farcs13 pixel$^{-1}$), which causes signals to not be Nyquist sampled. As a first step to alleviate undersampling, a dithering of the individual exposures was done during observations. Combining the dithering technique with upsampling during drizzling allows us to catch some of the sampled details of the PSF. As done in our previous IR WL analysis \citep{2017jee}, we let our calibration of galaxy shapes take care of the remaining undersampling bias.

\textbf{Non-linearity:} The response of the WFC3-IR detector follows a nearly linear relation until close to saturation where it then becomes nonlinear. Nonlinearity in the detector was reported at the 5\% level for saturated pixels \citep{2018dressel}. The \textit{calwf3} pipeline corrects the detector nonlinearity for pixels below the saturation level. As a precaution, we selected stars that are well below the saturation level when modeling the PSF.

\textbf{Brighter-fatter:} Analyzing the size-magnitude relation of the stellar frames that were used to model the PSF, we found a slope to the stellar locus with brighter objects tending to be larger. The brighter-fatter effect is well studied in CCDs and is thought to be caused by the electric field from the charges that have been accumulated in a pixel \citep{2014antilogus, 2015guyonnet}. For CCDs, \cite{2014antilogus} report that the size of the PSF increases by 2\% over the full dynamic range. However, few studies \citep{2017plazas, 2018plazas} have been carried out on the brighter-fatter effect in IR detectors.

The brighter-fatter effect requires attention for WL analyses because it will introduce a multiplicative bias to the measured shear. This is especially important for the faint galaxies that carry the WL distortion where forward-modeling an overly large PSF may lead to an overestimation of the shear. \cite{2015bmandelbaum} showed that a 1\% inflated PSF size introduces a systematic bias of m = 0.06 for a galaxy near the resolution limit. Our analysis of the stellar locus in the NGC104 frames shows that the average size of stars varies by as much as 5\% from the faintest detected objects to the saturation magnitude of the detector. In our PSF modeling, we intentionally avoid the stars near saturation. Thus, 5\% should be taken as an upper limit. Nevertheless, we desire to understand the systematic bias that might be introduced when forward-modeling a PSF with a size up to 5\% larger than the true PSF size. To do so, we simulated our forward-modeling shape measurement using Galsim\footnote{https://github.com/GalSim-developers/GalSim} \added{\edit1{\citep{2015rowe}}}. 

Simulated images of 10,000 S\'ersic profile galaxies (100 x 100 equally spaced on a grid) were created with the S\'ersic parameters sampled from the real galaxies of SpARCS1049. A uniform shear typical of a galaxy clusters ($\mytilde0.05$) was applied to the images. These simulated galaxy images were then convolved with a circular Gaussian PSF. Multiple passes of our shape-measurement pipeline were performed while forward modeling PSFs of size ranging from -15\% to +15\% of the true PSF size. This experiment showed that the multiplicative bias varies by m = 0.02 for a 5\% change in PSF size. At this level, the brighter-fatter effect has a low impact on galaxy cluster studies where shape noise is still the dominant uncertainty. However, in cosmic shear studies the brighter-fatter effect will need to be addressed.

\section{Weak-lensing method} \label{sec:wl_method}
\subsection{Theory}
At the core of weak gravitational lensing studies is the measurement of the minute distortion of galaxies. In the context of SpARCS1049, these distortions are caused by the altered light path that a photon travels while crossing the gravitational potential of the galaxy cluster. The altered light path can be described by its deflection angle - the angle between its original path away from its galaxy to its new path toward our telescope. The deflection angle is the gradient of the deflection potential. The differential transformation from the photon's emission position to the observed position is described by the Jacobian matrix:

\begin{equation}
    A = \left( \begin{array}{cc}
 1- \kappa - \gamma_1 & -\gamma_2 \\
 -\gamma_2 & 1 - \kappa + \gamma_1 \\
 \end{array}  \right) \label{eqn_A}
 \end{equation}
where the convergence $\kappa$ is an isotropic distortion defined as
\begin{equation} \label{eq:sigma_c}
    \kappa = \frac{\Sigma}{\Sigma_c}. 
\end{equation}
\noindent
In equation~\ref{eq:sigma_c}, $\Sigma$ is the projected mass density while $\Sigma_c$ is the WL critical surface density:
\begin{equation}
    \Sigma_c = \frac{c^2}{4\pi G D_l \beta}
\end{equation}
where c is the speed of light, G is the gravitational constant, $D_l$ is the angular diameter distance of the lens, and $\beta=D_{ls}/D_s$ is the lensing efficiency, which is the lens-source over source angular diameter distances. In equation~\ref{eqn_A}, the shear $\gamma$ is an anisotropic distortion and its two components can be combined to formulate the complex shear, $\gamma = \gamma_1 + i\gamma_2$. Observationally, the two distortion effects cannot be separated and the observed effect is the reduced shear $g_i = \gamma_i / (1 - \kappa)$.

Without the prior knowledge of the shape (ellipticity) of each galaxy image, measurement of $g$ directly based on a single galaxy image is not possible. Instead, the average complex ellipticity of an ensemble of galaxies is used to find $g$. This is done under the assumption that the average galaxy ellipticity is zero. We adopt the value of $\sigma_\text{int}=0.25$ for the intrinsic ellipticity dispersion; a value recently confirmed with the CANDELS data in \cite{2018schrabback}. This value of the intrinsic ellipticity dispersion is used in inverse-variance weighting when fitting models for mass measurement (Section \ref{sec:results}).

\subsection{Shape Measurement}
The WL observable, the reduced shear $g$, is ascertained by averaging the shapes of source galaxies. Our method of shape measurement is to fit a PSF-convolved elliptical Gaussian function to each object in the source catalog (source catalog defined in Section \ref{section:source_selection}). 

Postage stamp images of each object are cut from our F160W mosaic image. The size of each postage stamp image is chosen to be 12 times the semi-major axis of the object as determined by \texttt{Source Extractor}. This size reduces the effect of truncation bias that occurs when the light profile is prematurely truncated. However, a large postage stamp image increases the number of neighboring objects whose signal may contaminate the fit. We mask out the signal of the neighboring objects using the segmentation map output from \texttt{Source Extractor}. The difference between the light profile of the postage stamp image and the PSF-convolved elliptical Gaussian model is minimized with MPFIT \citep{2009markwardt}. We fix the centroid and background levels to the measurements from \texttt{Source Extractor} to reduce the free parameters of the fit. From the MPFIT output, we catalog the two complex ellipticity components 

\begin{align}
e_1 &= \frac{a-b}{a+b}\cos 2\phi, \\
e_2 &= \frac{a-b}{a+b}\sin 2\phi,
\end{align}
where $a$ and $b$ are the semi-major and -minor axes of the ellipse, respectively, and $\phi$ is the angle measured counter-clockwise from the positive x-axis. The ellipticity error $\sigma_e$ is also included into the catalog.

Measuring a galaxy's shape by fitting the light profile with an analytic function that does not perfectly represent the light profile introduces model bias. Moreover, the non-linear relation of the ellipticity measurement with the pixel noise causes noise bias. We correct for these biases by calibrating the ellipticities with a multiplicative factor \added{\edit1{of 1.25}} that is derived through simulations. Our method has been shown as effective by the sFIT method in the GREAT3 challenge \citep{2015mandelbaum}.

\subsection{Source Selection} \label{section:source_selection}
Selecting the source galaxies is an intricate step of a WL analysis. The lensing signal is observable only in the galaxies that are sufficiently behind the lens. Selection of source galaxies by spectroscopic or photometric redshift would be ideal but obtaining them is expensive and currently not possible with the limited \textit{HST} filter coverage. Instead, we select source galaxies based on their measured shape and photometric properties.

Galaxies residing in a cluster tend to be redder than field galaxies. The 4000\AA\ break, caused by the absorption of stellar light by ionized metals in stellar atmospheres, is a common feature in cluster galaxies and often gives rise to a red sequence in a color-magnitude diagram (CMD). For SpARCS1049, the 4000\AA\ break is redshifted to \mytilde 10,800\AA. This wavelength is encapsulated in the F105W and F160W filters. Figure 2 shows the CMD for SpARCS1049 with black dots representing the full object catalog. Cluster member galaxies selected from the Keck spectroscopic observations within $1.67 < z < 1.75$ and within the \textit{HST} imaging footprint are shown as red circles. These spectroscopic redshifts are detected from the $H\alpha$ emission line and give an active-galaxy selection bias to our cluster member sample. The BCG is shown as a red star and has a large magnitude separation from the other cluster members. The lack of a clear red sequence suggests that SpARCS1049 may be in an early state of formation.    

\begin{figure}[!ht]
    \centering
    \includegraphics[width=0.45\textwidth]{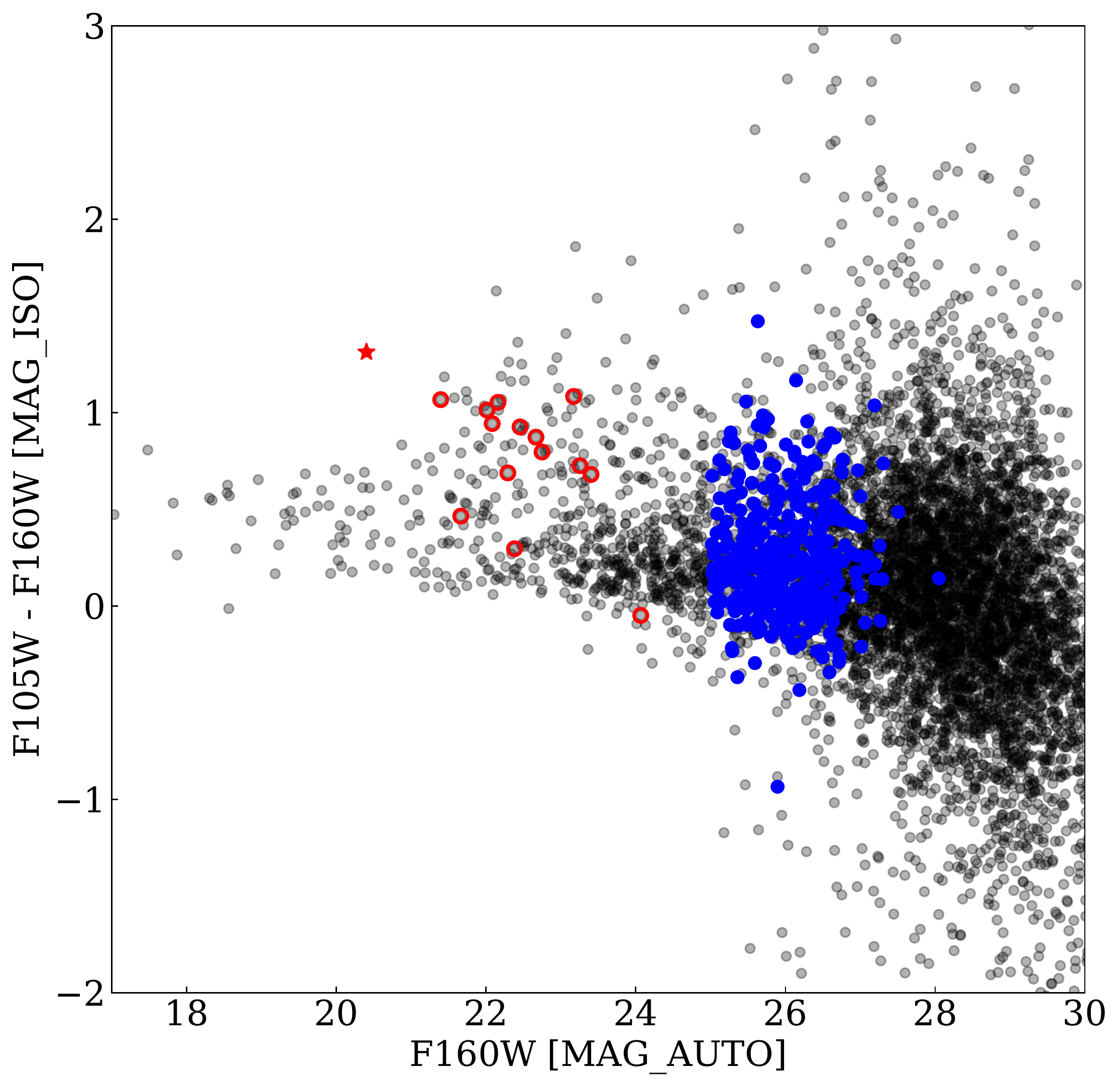}
    \caption{Color-magnitude diagram for SpARCS1049. Cluster member galaxies are marked red. The BCG is marked with a red star. Lensing source galaxies are depicted in blue. The bright limit of the source galaxies was chosen to maximize the lensing S/N and to mitigate contamination of cluster and foreground galaxies. The faint limit was set by requiring sources to have fitted ellipticity error $< 0.25$. }
    \label{fig:cmd}
\end{figure}
A pure source catalog is one that only contains lensed galaxies. Cluster member galaxies and foreground objects in the source catalog will contaminate the sample and dilute the lensing signal. Removal of these false sources is challenging without precise distances to each. Unfortunately, most removal techniques also filter out some true source galaxies. This is a problem because the lensing signal is proportional to the purity of the sample, whereas, the noise is proportional to $1/\sqrt{N}$. Furthermore, the uncertainty of the lensing efficiency $\beta$ increases with decreasing number of sources. Methods to maximize purity and source counts in the catalog vary. As a first step in defining a source catalog, we exclude foreground galaxies with an apparent magnitude cut that is fainter than the faintest spectroscopically confirmed cluster member. Our S/N tests show that retaining galaxies of F160W magnitude $>$ 25 provides the highest S/N. Including brighter galaxies decreases the detected WL signal and subsequently the S/N. 

In WL, sampling the faintest galaxies is desired because the most distant source galaxies are subject to the greatest lensing distortions. However, fitting a model to a low S/N galaxy is difficult and is subject to noise bias. To decrease noise bias, we exclude galaxies with a measured ellipticity error greater than 0.25. This constraint causes the faint magnitude limit seen in the CMD. In addition, galaxies in the source catalog are constrained to have a semi-minor axis greater than 0.3 pixels and ellipticity less than 0.9 to remove objects that are too small or too elongated to be galaxies. The total galaxy number density in our source catalog is \mytilde105 galaxies arcmin$^{-2}$.

\begin{figure}
    \centering
    \includegraphics[width=0.4\textwidth]{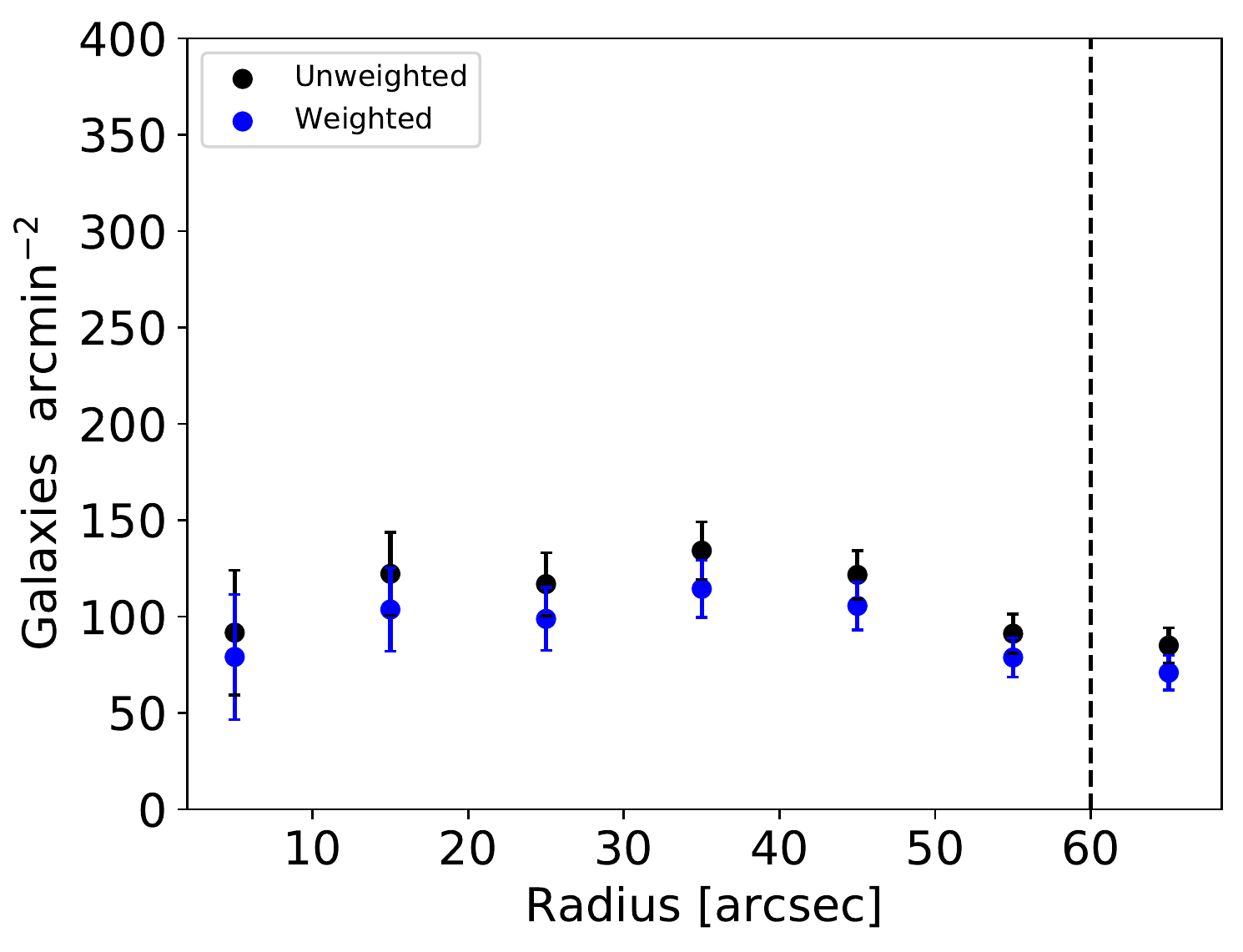}
    \caption{Black circles are radial number density of source galaxies centered on the BCG. Contamination by cluster galaxies might manifest as an overdensity near the cluster central region. The flat profile suggests that cluster member contamination is a minimum. The dashed vertical line represents the radius at which a circle is no longer complete in the \textit{HST} image. Error bars are Poissonian errors. \added{\edit1{Blue circles are the radial number density weighted by shape measurement ellipticity error 1 / ($\sigma_{int}^2$ + $\sigma_e^2$).}}}
    \label{fig:radial_density}
\end{figure}

To test the source catalog for contamination by cluster galaxies, we analyze the radial variation of source density. In Figure \ref{fig:radial_density}, the radial source density is shown with radial bins centered on the BCG. Contamination by cluster galaxies could manifest as an overdense region near the cluster center relative to the cluster outskirts. As seen in the figure, the radial number density of source galaxies is flat to $50\arcsec$. Beyond $50\arcsec$ the number density slightly decreases. This decrease is likely due to the limited frame coverage near the edge of the mosaic image and from the bright foreground galaxy drowning out background galaxies in the northern region of the image.

\subsection{Source Redshift Estimation}
%The WL signal is proportional to the lensing efficiency, $\beta$. For a given lens, this relation always increases with further source distance. However, the rate of increase in lensing efficiency for more distant sources is tied to the lens distance. High redshift lenses get smaller gains in lensing efficiency than low redshift lenses for more distant sources. This occurs because the angular diameter distance decreases beyond a redshift \mytilde1.5 due to its inclusion of the expansion of the Universe. Therefore, to achieve a high redshift weak-lensing analysis 

As shown in Equation \ref{eq:sigma_c}, the WL signal is proportional to the lensing efficiency, $\beta$. A proper characterization of $\beta$ relies on accurate knowledge of the angular diameter distances to the galaxy cluster and the source galaxies. However, the limited filter coverage for SpARCS1049 prevents direct calculation of distances to the source galaxies. As an alternative, we use the UVUDF photometric redshift catalog \citep{2015rafelski} as a control field, model it to represent our source catalog, and infer a representative distance to the source galaxies. 

\begin{figure}
    \centering
    \includegraphics[width=0.45\textwidth]{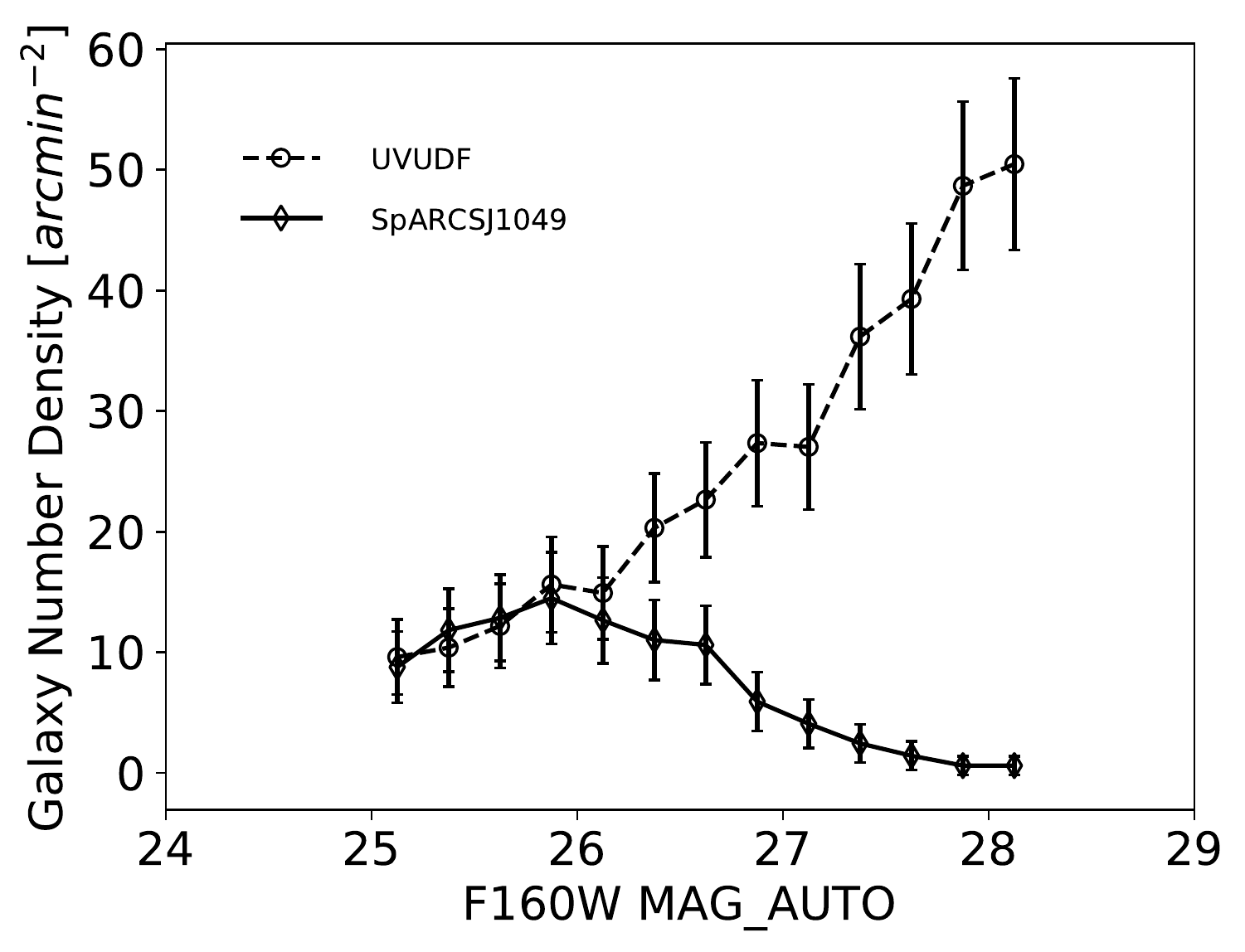}
    \caption{Magnitude distribution of galaxies in the SpARCS1049 WL source catalog. The UVUDF redshift catalog is used as a control field to estimate the redshift of the WL source catalog. In the 25 to 26 magnitude range the completeness in the SpARCS1049 and UVUDF catalogs are consistent. The discrepancy fainter than the 26th magnitude arises from the vastly different exposure times between our cluster imaging and the UVUDF. We compute the error bars assuming Poisson distributions.   }
    \label{fig:redshift}
\end{figure}

We constrain the UVUDF catalog with the same magnitude constraint specified in \textsection\ref{section:source_selection}. A comparison of the number density of galaxies in the source catalog and the UVUDF catalog is shown in Figure \ref{fig:redshift}. The number density of galaxies in the two catalogs is consistent in the 25 to 26 magnitude range. After the 26th magnitude the number density discrepancy can be attributed to the much deeper imaging of the UVUDF. To make the UVUDF catalog representative of our source catalog, we weight the UVUDF control catalog by the ratio of UVUDF to SpARCS1049 galaxy number density. The effective redshift and corresponding $\beta$ is calculated from the weighted UVUDF catalog as 

\begin{equation}
    \beta = \left< \mathrm{max} \left[ 0 , \frac{D_{ls}}{D_s} \right] \right>,
\end{equation}
where all foreground galaxies are assigned zero before averaging because they contain no lensing signal. From the weighted UVUDF catalog, we infer an effective redshift of 2.08 and $\beta = 0.107$ for our source catalog. Bias is introduced when representing all source galaxies by a single redshift. We reduce the bias as suggested in \cite{1997seitz} by taking the width of the beta distribution, $\left<\beta^2\right>=0.03$ into consideration. One may question whether the $\beta$ derived from a small field such as the UVUDF is representative of the small field of SpARCS1049. \cite{2014jee} compared the UVUDF to the UDF, GOODS-S, and GOODS-N redshift catalogs and found comparable $\beta$ values for each catalog. They reported the uncertainty of $\beta$ values between catalogs to affect mass estimates by at most \mytilde4\%. This small sample variance is attributed to the great depth of the {\it HST} image, which provides access to large distances along the line of sight.
Adding this uncertainty to the statistical uncertainty (\mytilde25\%) in quadrature shows that the statistical uncertainty on the mass will be dominant.

\section{Results} \label{sec:results}
\subsection{Mass Reconstruction}
A powerful aspect of WL is its ability to measure the projected mass distribution of the lens with minimal assumptions. There are multiple techniques that can be used to convert the observed shear $g$ to the convergence $\kappa$. We rely on the MAXENT method of \cite{2007bjee}, which converges to a solution that maximizes the entropy of a pixelized mass map while providing a reasonable goodness-of-the-fit for galaxy shapes.

\begin{figure*}[!ht]
    \centering
    \includegraphics[width=\textwidth]{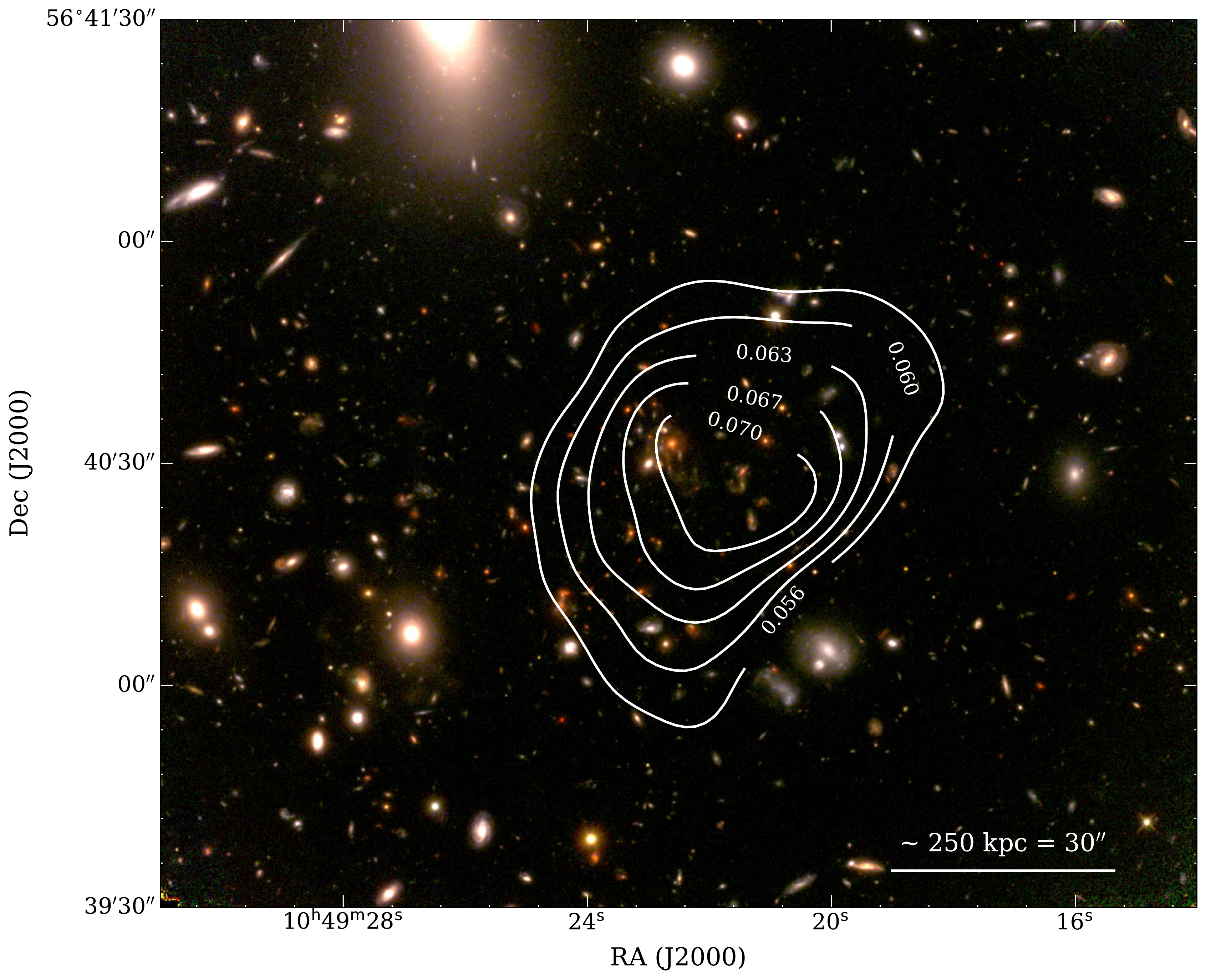}
    \caption{Mass reconstruction for SpARCS1049. Contour labels are subject to mass-sheet degeneracy. The distribution appears relaxed and does not show signs of substructures but does have a slight elongation in the east-west direction. The apparent offset of the mass peak and the BCG is shown to be insignificant through a bootstrap analysis. \added{\edit1{The significance (S/N) of the contours from the bootstrap result range from 2.0$\sigma$ for the lowest contour to 2.5$\sigma$ for the highest. The peak significance is 3.3$\sigma$. }}}
    \label{fig:convergence}
\end{figure*}

Figure \ref{fig:convergence} is the convergence map for SpARCS1049. The convergence is smoothed with a $\sigma=10\arcsec$ Gaussian kernel to remove a pixellation artifact. The convergence shows a slight elongation in the east-west direction but in general has a relaxed distribution \added{for the applied smoothing scale}. The mass peak lies \mytilde 10\arcsec (\mytilde90 kpc) to the southwest of the BCG. This offset, if significant, could be interpreted as an indication that the cluster mass is not centered at the BCG. To test the significance of the offset and the strength of our WL signal, we bootstrap the source catalog 1000 times. From the bootstrapped samples, we find that the cluster is detected at the 3.3$\sigma$ significance. The resampled catalogs also reveal that the $1\sigma$ uncertainty of the convergence peak location is \mytilde13\arcsec. Thus, we conclude that the mass map shows no statistically significant offset from the BCG.

\subsection{Mass Estimation}
Accurate estimation of the cluster mass is the primary goal of this work. There are numerous techniques that can be used to estimate the mass of a cluster from the observed galaxy ellipticity distribution. We choose to estimate the mass by fitting model profiles to the azimuthally averaged tangential shear. 

\begin{figure}
    \centering
    \includegraphics[width=0.45\textwidth]{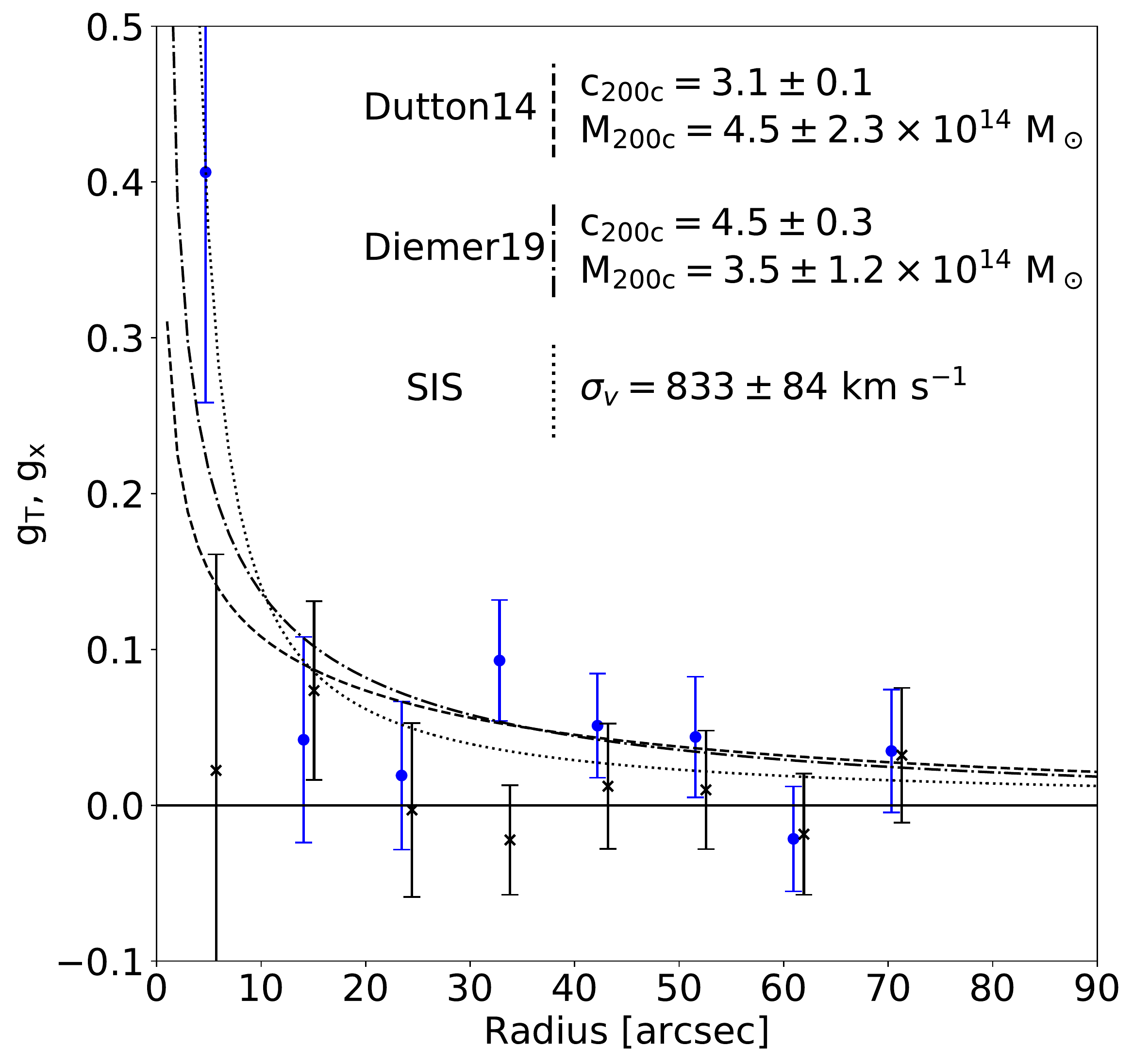}
    \caption{Density model fits to the tangential shear profile centered at the BCG. Blue circles are the tangential shear and black crosses are the cross shear. Error bars are the Poisson error. The cross shear has been shifted by $1\arcsec$ for display purposes. Three density profiles are shown: the SIS sphere, the $c$-$M$ relation of \cite{2014dutton}, and the $c$-$M$ relation of \cite{2019diemer}.}
    \label{fig:nfw_bcg}
\end{figure}

The reduced tangential shear $g_T$ at radius $r$ is a measure of the surface density contrast between the mean value within $r$ and the specific value at $r$
divided by $1-\kappa$. It is often written as the tangential components of the complex shear $g$

\begin{equation}
    g_T = -g_1\cos2\theta - g_2\sin2\theta
\end{equation}
where $\theta$ is the angle measured from the center of the cluster to the source, counter-clockwise. Rotating $\theta$ by 45 degrees gives the cross shear, which should be consistent with zero in the presence of no systematic effects and a circularly symmetric projected mass distribution. Figure \ref{fig:nfw_bcg} is the tangential shear measured in $10\arcsec$ bins centered on the BCG. The tangential shear profile is sensitive to the choice of center, particularly at small radii. We center our shear measurements on the BCG because it is consistent with the lensing peak and is an independent tracer of the cluster center. \added{\edit1{We also tested using the convergence peak as the center of the tangential shear fit and found that the derived mass is consistent with using the BCG as the center.}} The tangential shear profile clearly shows the detection of the lensing signal and the cross shear is consistent with zero. The outer limit of the tangential shear profile is set by the edge of the mosaic image and the data points beyond $60\arcsec$ are affected by the bright galaxy in the north. 

To estimate the mass, we fit 1D density models to the tangential shear as shown in Figure \ref{fig:nfw_bcg}. The first density profile that we fit is the singular isothermal sphere (SIS). The SIS profile returns a fitted velocity dispersion of $\sigma_v=833\pm84$ km s$^{-1}$.  Many density profiles have been derived from cosmological simulations that would all be appropriate to fit to the tangential shear. We fit some of the popular NFW-based models \added{\citep{1997navarro}} to our tangential shear so that direct comparison can be made with published galaxy cluster lensing masses. \added{\edit1{These fits are done by assuming the tangential shear profile follows a fixed concentration - mass (c-M) relation and fit only mass $M_{200c}$.}} \added{We utilize the \texttt{Colossus} code \citep{2018diemer} when performing the fits.} The masses are summarized in Table \ref{table:masses}. All three models return consistent masses. However, not all $c$-$M$ relations should be considered equal. As explained in detail in \cite{2019diemer}, the $c$-$M$ relation strongly depends on redshift and cosmology. Models that fit average concentration, such as \cite{2008duffy} and \cite{2014dutton}, are only valid under the assumed cosmology and redshift range of the simulations they are extracted from. Furthermore, power-law fits do not capture the upturn at high redshift and high mass as is shown in \cite{2015diemer}. \cite{2013ludlow} attribute the upturn to unrelaxed clusters. As SpARCS1049 is at a redshift of 1.71 and is likely in an early stage of formation, we suggest that the \cite{2019diemer} model is a good choice for a reasonably high mass cluster. Furthermore, of the three $c$-$M$ models that we fit, the \cite{2019diemer} model provides the best fit with reduced $\chi^2 = 1.03$. Throughout the discussion, we will use $3.5\pm1.2\times10^{14}$ M$_\odot$ for our WL mass estimation. 

\begin{table}[]
\centering
\caption{NFW density model fits to the tangential shear.}
\label{table:masses}
\def\arraystretch{1.5}
\begin{tabular}{l c c c}
\hline \hline
Model & $c_{200}$ & $M_{200c}$ $[\times10^{14} \text{M}_\odot]$ & $\chi_r^2$ \\ 
\hline
\cite{2008duffy} & $2.2\pm0.2$ & $5.9\pm3.5$ & 1.27\\
\cite{2014dutton} & $3.1\pm0.1$ & $4.5\pm2.3$ & 1.16 \\
\cite{2019diemer} & $4.5\pm0.3$ & $3.5\pm1.2$ & 1.03  \\
\hline \hline
\end{tabular}
\end{table}

%Out of interest, we also fit an NFW halo with both mass and concentration being simultaneously fit. Rather than fitting this model to the tangential shear, we fit the model shear to the ellipticity of each source galaxy. MCMC is used to determine the uncertainty in the two parameters (Figure ????). 

\begin{figure}
    
\end{figure}

\section{Discussion} \label{sec:discussion}
\subsection{Mass Comparison with Previous Studies}
A single previous study on the mass of SpARCS1049 exists. \cite{2015webb} estimated the mass of SpARCS1049 through a mass-richness scaling relation and by velocity dispersion of cluster member galaxies. They determined the abundance of cluster galaxies from Spitzer 3.6 $\mu$m observations to be $30\pm8$. This returned a mass of $\text{M}_{500\text{kpc}}=3.8\pm1.2\times 10^{14}\ \text{M}_{\odot}$ from application of the mass-richness scaling relation of \cite{2014andreon}. The authors note that the \mytilde30\% uncertainty on this mass does not take into consideration any redshift evolution of the scaling relation. 

To find the velocity dispersion of cluster galaxies, \cite{2015webb} obtained Keck spectroscopic measurements. The classification of cluster galaxies by these observations relied on the detection of the H$\alpha$ line. As the authors noted, this biases the sample to emission galaxies. Nevertheless, they classified 27 cluster member galaxies within 1500 km s$^{-1}$ of the mean cluster redshift and within 1.8 Mpc cluster-centric radius. The authors also mentioned that this included galaxies beyond the virial radius of the cluster. From the classified cluster members, the resulting velocity dispersion is $\sigma=430^{+80}_{-100}$ km s$^{-1}$ and the inferred mass is $\text{M}_\text{virial}=8 \pm 3 \times 10^{13}\ \text{M}_{\odot}$, after applying the velocity dispersion to virial mass relation of \cite{2008evrard}. The \mytilde40\% uncertainty reflects the unreliability of using strictly emission galaxies to derive the velocity dispersion. It is peculiar that the mass from velocity dispersion is much lower than from the mass-richness relation. It goes against the notion that emission galaxies should be infalling and have inflated velocity dispersion.

\begin{figure}
    \centering
    \includegraphics[width=0.45\textwidth]{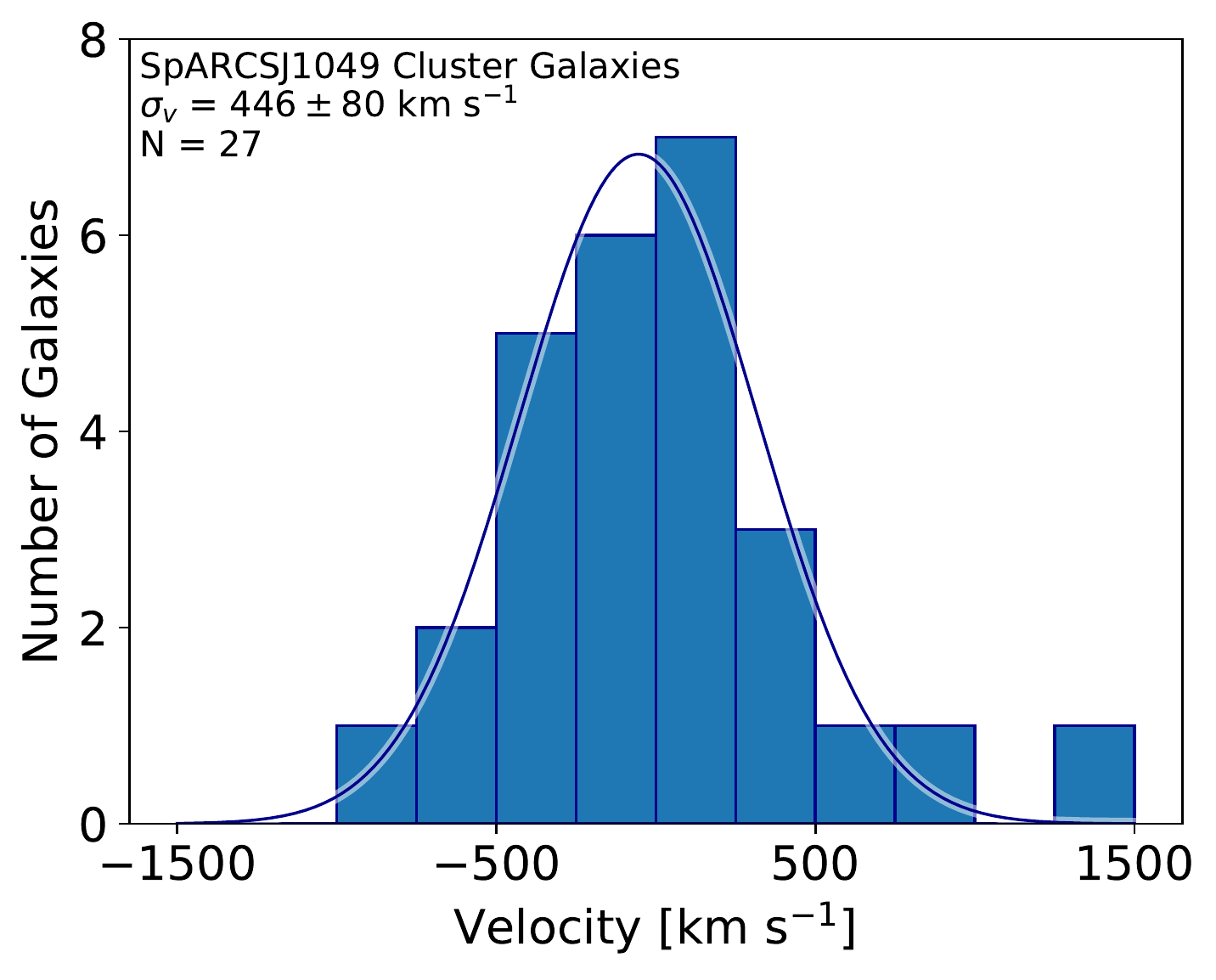}
    \caption{Velocity histogram of cluster member galaxies selected within 1500 km s$^{-1}$ of the average velocity. The blue line is the best-fit Gaussian model. An Anderson-darling test fails to reject the null hypothesis that the galaxies follow a normal distribution.}
    \label{fig:vel_disp}
\end{figure}

Using an updated spectroscopic redshift catalog, we selected member galaxies in the same manner as \cite{2015webb}. Applying the bi-weight velocity dispersion to the 27 detected members gives $\sigma_v=446\pm80$ km s$^{-1}$ and $\text{M}_\text{virial}=1.0\pm0.4\times10^{14}$ M$_\odot$ with the conversion from \cite{2008evrard}. We attach the same 40\% uncertainty on mass as \cite{2015webb}. This mass is consistent with the findings of \cite{2015webb}. Figure \ref{fig:vel_disp} shows the histogram of cluster galaxies. Performing an Anderson-darling test on the cluster galaxies fails to reject the null hypothesis that they follow a normal distribution. 

Our WL mass estimate $\text{M}_{200} = 3.5\pm1.2\times10^{14}\ \text{M}_\odot$ provides the first mass estimate free of a dynamical equilibrium assumption. This mass estimate is consistent with the mass-richness estimation. However, there is a discrepancy with the mass from the velocity dispersion.

\subsection{Rarity}
Massive galaxy clusters at high redshift are expected to be rare according to the hierarchical structure formation model. SpARCS1049 was selected for this study because of its known large mass and should be tested for its rarity. Future work will fully analyze the rarity of the See Change sample of massive galaxy clusters between redshift 1.10 and 1.75.

We determine the rarity of this cluster by integrating the number of clusters above a minimum mass and redshift as

\begin{equation}
    N(M,z) = \int^\infty_{z_{\text{min}}}\int^\infty_{M_{\text{min}}} \frac{dV(z)}{dz} \frac{dn}{dM}dM dz
\end{equation}
where $dV/dz$ is the volume element and $dn/dM$ is the mass function. We set the lower limits of the integrals to $z_{\text{min}}=1.71$ and $M_{\text{min}}=3.5\times10^{14}\ \text{M}_\odot$, the central mass estimate. The exact upper limits of the integral are insignificant because the rarity of the cluster (steepness of the mass function in this regime) causes the integral to converge quickly.
Using HMFCalc \citep{2013murray}, we adopt the mass function of \cite{2008tinker} that has been updated by \cite{2013behroozi}. The estimated abundance of a cluster with mass and redshift of SpARCS1049 is \mytilde12 over the full sky or \mytilde0.01 clusters within the \mytilde41.9 deg$^2$ footprint of SpARCS. Alternatively, taking the 1-$\sigma$ lower limit of our mass estimation result $M_{\text{min}}=2.3\times10^{14}\ \text{M}_\odot$ gives a rarity of \mytilde185 clusters in the entire sky or \mytilde0.2 in the SpARCS field. For comparison, the rarity of two additional See Change clusters, IDCS J1426+3508 (z=1.75) and SPT-CL J2040-4451 (z=1.48), are \mytilde1200 and \mytilde1 clusters in the full sky, respectively, using their WL measured central mass values \citep{2017jee}. Thus, SpARCS1049 is similar in rarity to other See Change clusters.

This type of rarity calculation has well-documented limitations \citep[][]{2011hotchkiss, 2012hoyle, 2013harrison}. As pointed out by \cite{2011hotchkiss}, the rarity integral only considers clusters that have mass and redshift greater than or equal to the selected lower limits. The rarity calculation neglects equally rare clusters that exist at higher mass but lower redshift and vice versa, which results in a bias that causes low rarities. Furthermore, the rarity calculation relies on integration of a mass function that is derived from cosmological simulations that often poorly reproduce the high-mass high-redshift end of the mass function. \cite{2013murrayb} report that the halo mass function has \mytilde20\% uncertainty at the high mass end. An additional limitation comes from Eddington bias \citep{1913eddington,2011mortonson}. Eddington bias occurs because the mass function of the universe is steeply declining with increasing mass at the mass and redshift of SpARCS1049. Therefore, it is more likely to overestimate a cluster mass than to underestimate a cluster mass for such an extreme object.

\section{Conclusions} \label{sec:conclusion}
An \textit{HST}-IR WL analysis of the massive galaxy cluster SpARCS1049 is presented. \textit{HST}-IR detector systematics have been quantified with a specific focus on the brighter-fatter effect. Our simulations show that the brighter-fatter effect gives at most a 2\% shape bias in our shear measurements. The systematics discussed will be important for future WL studies with next generation telescopes, such as JWST, Euclid, and WFIRST. 

The projected mass distribution has been reconstructed from the averaged background galaxy ellipticities. The mass distribution is seemingly relaxed \added{for the applied smoothing scale} with the centroid consistent with the BCG. We have found the mass of the cluster to be $3.5\pm1.2\times10^{14}\ \text{M}_\odot$ for our best-fit NFW model. This mass is consistent with the mass estimated from a mass-richness scaling relation. However, it is inconsistent with the mass from velocity dispersion of spectroscopically confirmed cluster galaxies. Finally, we have tested the mass of the cluster for its rarity. We have found the expected abundance of similarly massive clusters to be $<1$ within the parent survey, thus suggesting that SpARCS1049 is a uniquely massive cluster.

\acknowledgments
Support for the current {\it HST}~program was provided by NASA through a grant from the Space Telescope Science
Institute, which is operated by the Association of Universities for Research in Astronomy, Incorporated, under
NASA contract NAS5-26555.
This study is supported by the program Yonsei University Future-Leading Research Initiative.
M. J. Jee acknowledges support for the current research from the National Research Foundation of Korea under the programs 2017R1A2B2004644 and 2017R1A4A1015178. 
GW acknowledges support from the National Science Foundation through grant AST-1517863, by HST program number GO-15294, and by grant number 80NSSC17K0019 issued through the NASA Astrophysics Data Analysis Program (ADAP).

\bibliography{mybib}{}
\bibliographystyle{aasjournal.bst}

%\listofchanges
\end{document}